\documentclass{acmsiggraph}

\title{
  Plausible Shading Decomposition For Layered Photo Retouching
}

\author{
Carlo Innamorati\qquad
Tobias Ritschel\qquad
Tim Weyrich\qquad
Niloy Mitra
\\
University College London
}
\pdfauthor{}


\setcopyright{none}

\copyrightyear{2016}

\conferenceinfo{SIGGRAPH 2016}{July 24-28, 2016, Anaheim, CA}
\isbn{978-1-4503-ABCD-E/16/07}
\doi{http://doi.acm.org/10.1145/9999997.9999999}

\usepackage{amsmath}
\usepackage{paralist}
\usepackage{booktabs}
\usepackage{microtype}
\usepackage{amssymb}
\usepackage{soul}
\usepackage[usenames,dvipsnames]{xcolor}
\usepackage[normalem]{ulem}
\usepackage{multirow}
\usepackage{tikz}
\usepackage{tabularx}
\usepackage{units}
\usepackage{mathptmx}
\usepackage{mathrsfs}
\usepackage{wrapfig}

\newcolumntype{C}[1]{>{\centering}m{#1}}

\newcommand{\eg}{e.\,g.,\ }
\newcommand{\ie}{i.\,e.,\ }
\newcommand{\etal}{~et~al.\ }

\newcommand{\rpm}{\raisebox{.2ex}{$\scriptstyle\pm$}}

\DeclareGraphicsExtensions{.pdf,.ai,.eps,.jpg}
\def\figurePath{Figures/}

\def\myfigure#1#2{\begin{figure}[h]\centering\includegraphics*[width = \linewidth]{\figurePath#1}\caption{#2}\label{fig:#1}\end{figure}}

\def\mytfigure#1#2{\begin{figure*}[t!]\centering\includegraphics*[clip, width = \linewidth]{\figurePath#1}\caption{#2}\label{fig:#1}\end{figure*}}

\def\mysection#1#2{\section{#1}\label{sec:#2}}
\def\mysubsection#1#2{\subsection{#1}\label{sec:#2}}

\newcommand{\refSec}[1]{Sec.~\ref{sec:#1}}
\newcommand{\refFig}[1]{Fig.~\ref{fig:#1}}
\newcommand{\refEq}[1]{Eq.~\ref{eq:#1}}

\def\clap#1{\hbox to 0pt{\hss #1\hss}}%
\def\initials#1{\protect\clap{\smash{\raisebox{1.4ex}{\tiny{\textsf{\textit{#1}}}}}}}%
\makeatletter
\newcommand{\EDIT}[4][]{\protect\@ifundefined{hidecomments}{%
  \strut{\color{#3}{\hspace{0pt}\initials{#2}\protect\sout{#1}{#4}}}%
  }{}}
\newcommand{\NOTEboxed}[3]{\protect\@ifundefined{hidecomments}{%
  {\centering\fbox{\parbox{0.97\linewidth}{\protect\EDIT{#1}{#2}{#3}}}}
  }{}}
\makeatother
\def\ignore#1{}

\usepackage{bm}
\definecolor{teal}{rgb}{0, .5, .5}

\begin{document}

\teaser{
  \includegraphics[width=\linewidth]{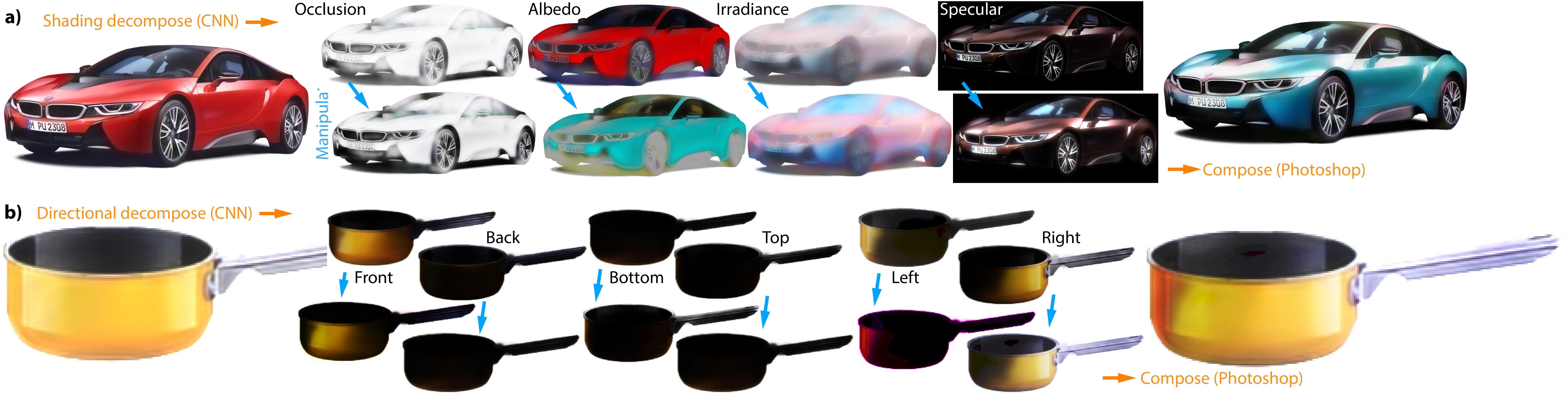}
  \caption{
  Our approach automatically splits input images into layers motivated by light transport, such as \emph{(a)}: occlusion, albedo, irradiance and specular, or \emph{(b)}: the six major spatial light directions, which can then be manipulated independently using off-the-shelf photo manipulation software and composed back to an improved image.
  For \emph{(a)} shadows were made deeper, albedo hue changed,  saturation of irradiance increased and the specular was blurred for a more glossy material.
  For \emph{(b)} The front lighting was made weaker and light from the left had been tinted red.}
    \label{fig:Teaser}
}

\maketitle

\begin{abstract}
Photographers routinely compose multiple manipulated photos of the same scene (layers) into a single image, which is better than any individual photo could be alone.
Similarly, 3D artists set up rendering systems to produce layered images to contain only individual aspects of the light transport, which are composed into the final result in post-production.
Regrettably, both approaches either take considerable time to capture, or remain limited to synthetic scenes.
In this paper, we suggest a system to allow decomposing a single image
into a \emph{plausible shading decomposition} (PSD) that approximates
effects such as shadow, diffuse illumination, albedo, and specular
shading.
This decomposition can then be manipulated in any off-the-shelf image manipulation software and recomposited back.
We do so by learning a convolutional neural network trained using synthetic data.
We demonstrate the effectiveness of our decomposition on synthetic (i.e., rendered) and real data (i.e., photographs), and use them for common photo manipulation, which are nearly impossible to perform otherwise from single images.
\end{abstract}



\mysection{Introduction}{Introduction}
Professional photographers regularly compose multiple photos of the same scene into one image, giving themselves more flexibility and artistic freedom than achievable by capturing in a single photo.
They do so, by `decomposing' the scene into individual \emph{layers}, \eg by changing the light, manipulating the individual layers (\eg typically using a software such as Adobe Photoshop), and then composing  them into a single image.
On other occasions this process is called stacking. 
Unfortunately, this process requires the effort of setting up and taking multiple images. 
An alternative that overcomes this limitation is rendering synthetic images.
In this case, the image can be clearly decomposed into the individual aspects of light transports (\eg specular highlights vs.\ diffuse shading).
The light path notation \cite{heckbert1990adaptive} provides a strict criterion for this decomposition.
Regrettably, this requires the image to be ``synthesizable'', \ie material, geometry, and illumination should be known as well as a suitable simulation algorithm.
This is often not the case for scenes obtained as single images.

In this work, we set out to devise a system that combines the strength of both approaches: the ability to work on real photos, combined with a separation into light transport layers. 
Starting from a single photograph, our system produces a decomposition into layers, which can then be individually manipulated and recombined into the desired image using off-the-shelf image  manipulation software. \refFig{Teaser} shows an example.

While many decompositions are possible, we suggest a specific layering model that is inspired by how many artists as well as practical contemporary rendering systems (\eg in interactive applications such as computer games) work: a decomposition into shadow, diffuse illumination, albedo, and specular shading.
This model is not completely physical, but simple, intuitive for artists and its inverse model is effectively learnable.
We formulate shadow as a single scalar factor to brighten or darken the appearance, resulting from adding the diffuse and specular shading.
The diffuse shading is further decomposed into illumination (color of the light) and reflectance (color of the object), while the specular shading is modeled directly.
To invert this model, we employ a deep convolutional architecture (CNN) that is trained using synthetic data, for which the ground truth-decomposition of a photo into light transport layers is known.

In summary, we make the following contributions: 
\begin{itemize}
\item splitting and re-combination of images based on light transport layers (shadow, diffuse light, albedo and specular shading); 
\item a CNN trained on synthetic data to perform such a split; and 
\item evaluating our approach on a range of real photographs and demonstrating utility for photo-manipulations. 
\end{itemize}

\mysection{Previous Work}{PreviousWork}
Combining multiple photos (also referred to as a ``stack'') of a scene where one aspect has changed in each layer is routinely used in computer graphics \cite{cohen2003image}. For example, NVIDIA IRay actively supports rendered LPE layers (light path expressions \cite{heckbert1990adaptive}) to be individually edited to simplify post-processing towards  artistic effects without resorting to solving the inverse rendering problem. 
One aspect to change is illumination, such as flash-no-flash \cite{eisemann2004flash} or exposure levels \cite{mertens2009exposure}.
More advanced effect involve direction of light~\cite{akers2003conveying,Rusinkiewicz:2006:ESF,fattal2007multiscale}, eventually resulting in a more sophisticated user interface~\cite{boyadzhiev2013user}.
All these approaches require specialized capture to gather multiple images captured by making invasive changes to the scene, limiting their use in practice to change an image post-capture. In fact, several websites and dedicated YouTube channels have emerged to provide DIY instructions to setup such studio configurations.

For single images, a more classic approach is to perform intrinsic decomposition into shading (irradiance) and diffuse reflectance (albedo) \cite{barrow1978recovering,garces2012intrinsic,bell2014intrinsic}, possibly supported by a dedicated UI for images \cite{bousseau2009user,boyadzhiev2012user}, using annotated data~\cite{bell2014intrinsic}, or videos \cite{ye2014intrinsic,bonneel2014interactive}.
Recently, CNNs have been successfully applied to this task  producing state-of-the-art results~\cite{narihira2015direct}.

We also using a data-driven CNN-based approach to go beyond classic intrinsic image decomposition light transport layers with further separation into occlusion and specular components, that are routinely used when post-compositing layered renderings (see \refSec{Results} and supplementary materials).

In other related efforts, researchers have looked into factorizing components, such as specular~\cite{tan2004separating,mallick2006specularity} from single images, or ambiant occlusion~(AO) from single~\cite{yang2015ambient} or multiple captures~\cite{hauagge2013photometric}. 
We show that our approach can solve this problem at a comparable quality, but requires only a single photo and in combination yields further separation of diffuse shading and albedo without requiring a specialized method.

Despite the advances in recovering reflectance (\eg with two captures and a stationarity assumption \cite{aittala2015two}, or with dedicated UIs \cite{dong2011appgen}), illumination (\eg Lalonde\etal\shortcite{lalonde2009estimating} estimate sky environment maps and  Rematas\etal\shortcite{RematasCVPR2016} reflectance maps) and depth (\eg Eigen\etal\shortcite{eigen2014depth} use a CNN to estimate depth) from photographs, no system doing a practical joint decomposition is known. Most relevant to our effort, is SIRFS~\cite{BarronTPAMI2015} that build data-driven priors for shape, reflectance, illumination, and use them in an optimization setup to recover the most likely shape, reflectance, and illumination
under these priors (see \refSec{Results} for explicit comparison to SIRFS). 

In the context of image manipulations, specialized solutions exist: 
Oh\etal\shortcite{oh2001image} represent a scene  as a layered collection of color and depth to enable distortion-free copying of parts of a photograph, and allow discounting effect of illumination on uniformly textured areas using bilateral filtering; 
Khan\etal\shortcite{khan2006image} enable automatically replacing one material with another (\eg increase/decrease specularity, transparency, etc.) starting from a single high dynamic range image by exploiting our `blindness' to certain physical inaccuracies; 
Caroll\etal\shortcite{carroll2011illumination} achieve consistent manipulation of inter-reflections; or the system of Karsch\etal 
\shortcite{karsch2011rendering} that combines many of the above into state-of-the art and compelling augmented image synthesis.

Splitting into light path layers is typical in rendering inspired by the classic light path notation \cite{heckbert1990adaptive}.
In this work, different from Heckbbert's physical $E(S|D)^*L$ formalism, we use a more edit-friendly factorization into shadow, diffuse light, diffuse material, and specular, instead of separating direct and indirect effects.
While all the above works on photos, it was acknowledged that rendering beyond the laws of physics can be useful to achieve different artistic goals~\cite{todo2007locally,vergne2009light,ritschel2010interactive,LayeredVectorisation,dong2015measurement,schmidt2015state}.
Our approach naturally supports this option, allowing users to freely change light transport layers, using any image-level software of their choice, also beyond what is physically correct. 
For example, the StyLit system proposed by Fi\v{s}ser\etal\shortcite{fiser2016stylit} correlates artistic style with light transport expressions. Specifically, it requires pixels in the image to be labeled with light path information, \eg by rendering and aligning. Hence, it can take the output of our factorization to enable stylization single photographs without being restricted to rendered content.

\mysection{Our Approach}{OurApproach}

\paragraph{Overview.}
Our system has three main components:
(i)~producing training data (\refSec{TrainingData}); 
(ii)~a convolutional neural network to decompose single images into light transport layers (\refSec{Decomposition}); and
an interactive system to manipulate the light transport layers before recomposing them into an image (\refSec{Composition}).

The training data (\refSec{TrainingData}) is produced by rendering a large number of 3D scenes into image tuples, where the first is the composed image, while the other images are the light transport layers.
This step needs only to be performed once and the training data will be made available upon publication.

The layer decomposition (\refSec{Decomposition}) is done using a CNN that consumes a photo and outputs all its light transport layers.
This CNN is trained using the training data from the previous step.
We selected a  convolution-deconvolution architecture that is only to be trained once, can be executed efficiently on new input images, and its definition will be made publicly available upon publication (please refer to the supplementary for the architecture).

Optionally, we employ an upsampling step \refSec{Upsampling}, that re-samples the fixed-resolution light transport layer CNN output, such that composing them in the arbitrarily high resolution of the original resolution produces precisely the original image without any bias, blur, or drift.

Finally, we suggest a system (\refSec{Composition}) that executes the CNN on a photo at deployment time to produce light transport layers, which can then be individually and interactively manipulated, in any off-the-shelf image manipulation software, allowing operations on photos that previously were only possible on layered renderings, or using multiple captures.  

Before detailing all the three steps, we will next introduce the specific image formation model underlying our framework in \refSec{Model}.

\mysubsection{Model}{Model}
We propose two different image formation models: one that is invariant under light direction and one that captures the directional dependency.

\myfigure{Model}{The components of our two imaging models.}

\paragraph{Non-directional Model.}
We model the color $C$ of a pixel as 
\begin{align}
\label{eq:Model}
C = O  (\rho  I  + S), 
\end{align}
where $O\in(0,1)\in\mathbb R$ denotes the \emph{occlusion}, which is the fraction of directions in the upper hemisphere that is blocked from the light;  
the variable $I\in(0,1)^3\in\mathbb R^3$ denotes the \emph{diffuse illumination} (irradadiance), \ie color of the light, without any directional dependence;  $\rho\in(0,1)^3\in\mathbb R^3$ describes the \emph{albedo} (diffuse reflectance), \ie the color of the surface itself; 
and finally, $S\in(0,1)^3\in\mathbb R^3$ is the \emph{specular shading}, where we do not separate between the reflectance and the illumination, and do not capture any directional dependence.

\paragraph{Directional Model.}
The direcitonal model is
\begin{align}
C
=
\sum_{i=1}^N
\mathbf R (\int_\Omega
L_\mathrm i(\omega)
b_i(\omega)
\mathrm d \omega
),
\end{align}
where $L_\mathrm i$ is the incoming light, and $\mathbf R$ the reflection operator, mapping incoming to outgoing light.
In other words, we express pixel color as a sum of reflections of $n$ basis illuminations.

Here, $(b_1,\ldots,b_n)$ can be any set of spherical functions that sum to 1 at every direction, \ie $\sum_i^n b_i(\omega)=1$ (partition of unity).
One such decomposition is the spherical harmonics basis of any order or the cube basis (that is one for a single cube face).
In our approach, we suggest to use a novel \emph{soft cube} decomposition that combines strengths of both: It if very selective in the directional domain, has finitely many components but also does not introduce a sharp cut in the directional domain.
It is defined as the clamped dot product between the $i$-th cube side direction $\mathbf c_i$ raised to a sharpening power $\sigma=20$:
$$
b_i(\omega) = 
\max(\left<\omega,\mathbf c_i\right>, 0)^\sigma
/
\sum_{j=1}^6
\max(\left<\omega,\mathbf c_j\right>, 0)^\sigma. 
$$
An additional benefit of the (soft) cube decomposition is, that it is, other than SH, is strictly positive, facilitating loading of layers into applications that do not (well) support negative values, such as Photoshop.
Other bases are possible in this framework, allowing to tailor it to specific domain, where a prior on light directions might exist (\eg portrait photos). 

There are many values of $O$, $I$, $\rho$ and $S$ to explain an observed color $C$, so the decomposition is not unique.
Inverting this mapping from a single observation is likely to be  impossible.
At the same time, humans are clearly able to solve this task.
One explanation can be that they rely on context, on the spatial statistics of multiple observations $c(\mathbf x)$, such that a decomposition into light transport layers becomes possible. In other words, simply not all arrangements of decompositions are equally likely.
As described next, we  employ a CNN to learn this decomposition in a similar fashion.

\myfigure{TrainingData}{Samples from our set of synthetic training data.}
\mysubsection{Training data}{TrainingData}
Training data comprises of synthetic images that show a random shape, with partially random reflectance shaded by random environment map illumination.

\paragraph{Shape.} 
Shape geometry comprises of 300 random cars from from ShapeNet~\cite{ChangFGHHLSSSSX15}. Note that the models were assumed to be upright. 
This class was chosen, as it presents both smooth surfaces as well as hard edges typical for mechanical objects.
Note that our results show many classes very different from cars, such as fruits, statues, mechanical appliances, etc.
Please note that we specifically restricted training to only cars to evaluate how the CNN generalizes to other object classes. 
Other problems like optical flow have been solved using CNNs on general scenes despite being trained on  very limited geometry, such as training exclusively on chairs \cite{dosovitskiy2015flownet}.

\paragraph{Reflectance.}
Reflectance using the physically-corrected Phong model \cite{lafortune1994using}, sampled as follows:
The diffuse colors come directly from ShapeNet models.
The specular component $k_\mathrm s$ is assumed to be a single color.
A random decision is made if the material is assumed to be electric or dielectric.
If it is electric, we choose the specular color to be the average color of the diffuse texture.
Otherwise, we choose it to be a uniform random grey value.
Glossiness is set as $n=3.0^{10\xi}$, where $\xi$ is a random value  in $U[0,1]$. 

\paragraph{Illumination.}
Illumination is sampled from a set of 122 HDR environment maps in resolution $512\times 256$ that have an uncalibrated absolute range of values but are representative for typical lighting settings: indoor, outdoor, as well as studio lights.

\paragraph{Rendering.}
After fixing shape, material, and illumination, we synthesize a single image from a random view (rotation is only about vertical).
To compute $C$, we compute all components individually, and compose them according to \refEq{Model}.
The occlusion term $O$ is computed using screen-space occlusion \cite{ritschel2009approximating}.
The diffuse shading $I$ is computed using pre-computed irradiance environment maps \cite{ramamoorthi2001efficient}.
Similarly, specular shading is the product of the specular color $k_\mathrm s$ selected according to the above protocol, and a pre-convolved illumination map for gloss level $n$.
Diffuse albedo $k_\mathrm s$ is directly available in ShapeNet.
While we could also try to infer the glossiness, it would not be clear how to use its non-linear effect with classic layering.
No indirect illumination or local interactions are rendered.
When learning the directional-dependent variant, we render six images, where the illumination was pre-convolved with the $i$-th decomposed illumination.

While this image synthesis is far from being physically accurate, it can be produced easily, systematically and for a very large number of images, making it suitable for learning the layer statistics.
Overall we produce 100\,k images in a resolution of $256\times 256$ (ca.\ 10\,GB) in 5 hours on a current PC with a decent GPU.

\paragraph{Units.}
Care has to be taken in what color space learned and training data is to be processed.
As the illumination is HDR, the resulting image is an HDR rendering.
However, as our input images will be LDR at deployment time, we need to match their range.
To this end, automatic exposure control is used to map those values into the LDR range, by selecting the $0.95$ luminance percentile of a random subset of the pixels and scale all values such that this value maps to $1$.
The rendered result $C$ is stored after gamma-correction with $\gamma=2.0$.
All other components are stored in physically linear units ($\gamma=1.0$) and are processed in physical linear units by the CNN and the end-application using the layers.
Doing the final gamma-correction will consequentially be up to the application using the layers later on (as shown in our edit examples).

\mytfigure{DirectionalDecomposition}{Decomposition of input images (left) into the six  directional layers (left) for different objects.}

\mysubsection{Decomposition}{Decomposition}

We perform decomposition  using a CNN \cite{krizhevsky2012imagenet} trained using the data produced as described above.
Input to the network is a single image such as a photograph.
Output for the non-directional variant are the four images (occlusion, diffuse illumination, albedo, specular shading), the light transport layers, where occlusion is scalar and the others are three-vector-valued.
Output for the directional variant, the output are 6 directional diffuse and specular layers, to be composed with one AO, and one albedo layer.

This design follows the convolution-deconvolution with crosslinks idea, resulting in an hourglass scheme \cite{ronneberger2015u}.
The network is fully-convolutional. We start at a resolution of $256\times 256$ that is reduced down to $2\times 2$ through stride-two convolutions. We then perform two stride-one convolutions to increase the number of feature layers in accordance to the required number of output layers (i.e. quadruple for the light transport layers, sextuple for the directional light layers). The deconvolution part of the network comprises of blocks performing a resize-convolution (upsampling followed by a stride-one convolution), crosslinking and a stride-one convolution. Every convolution in the network is followed by a ReLU \cite{nair2010relu} non-linearity except for the last layer, for which a Sigmoid non-linearity is used instead. This is done to normalize the output to the range $[0, 1]$.
Images larger or smaller than the required input size of $256\times 256$ will be appropriately scaled and/or padded to be square with white pixels.
All receptive fields are $3\times 3$ pixels in size except for the first and last two layers that are $5\times 5$.

As the loss function, we combine a per-light transport layer L2 loss with a novel three-fold \emph{recombination} loss, that encourages the network to produce combinations that result in the input image and fulfils the following requirements:
(i)~the layers have to produce the input, so $C=AO(I\rho+S)$; 
(ii)~the components should explain the image without AO, \ie $C/AO=I\rho+S$; and 
(iii)~diffuse reflected light should explain the image without AO and specular, so $C/AO-S=I\rho$.
If the network was able to always perform a perfect decomposition, the L2 loss alone would be sufficient.
As it makes errors in practice, the second loss biases those errors to at least happen in such a way that the combined result does not deviate from the input.
All losses are in the same RGB-difference range and were weighted equally for simplicity.

Overall, the network is a rather straight-forward modern design, but trained to solve a novel task (light transport layers) on novel kind of training data (synthesized, directionally-dependant information).
We used TensorFlow~\cite{ChangFGHHLSSSSX15} for our implementation platform and each model requires only several hours to train (both have been trained for 12 hours). A more detailed description of the network's architecture can be found in the supplemental materials.

\mysubsection{Upsampling}{Upsampling}
Upsampling (\refFig{Upsampling}) is an optional step applied to input images that have a resolution arbitrarily higher than the one the CNN is trained on (256$\times$256).
Input to this process are all the layers in the low resolution.
Output are the layers in that arbitrarily high resolution, such that applying the composition equation results in the high-resolution image.

\myfigure{Upsampling}{Our CNN computes a decomposition in a fixed resolution of $256\times256$ with the results \emph{(top insets)}.
Given the HD original, we perform upsampling \emph{(bottom insets)} that assures they combine to the HD input when blended.
}

This is achieved as follows, independently for every high-resolution pixel and in 100 iterations per pixel:
Initially all value are set to the CNN output.
At each iteration, we hold all layer components fixed, and in turn solve for the missing one, given the color.
We then blend this result gradually (weight 0.001) with the previous result.
This is repeated for all layers.
Additionally, the light layer is forced to not change the chroma.
When values leave the unit RGB cube, they are back-projected.
The result is a layering in an arbitrary resolution that follows the CNN decomposition, yet produces the high resolution image precisely (energy-conserving).
An immediate practical consequence of this is, that any image loaded into our system after the decomposition into layers looks precisely like the input without any initial bias (blurr or color shift) introduced by the CNN processing. Please note that we explicitly mention upsampling for the results where we use this mode (only for edits). 

\myfigure{Edits}{$O\cdot (\rho \cdot I + S)$ editing (See text ``Edit'' \refSec{Results}).
}

\begin{figure}[t!]
    \centering
    \includegraphics[width=.95\columnwidth]{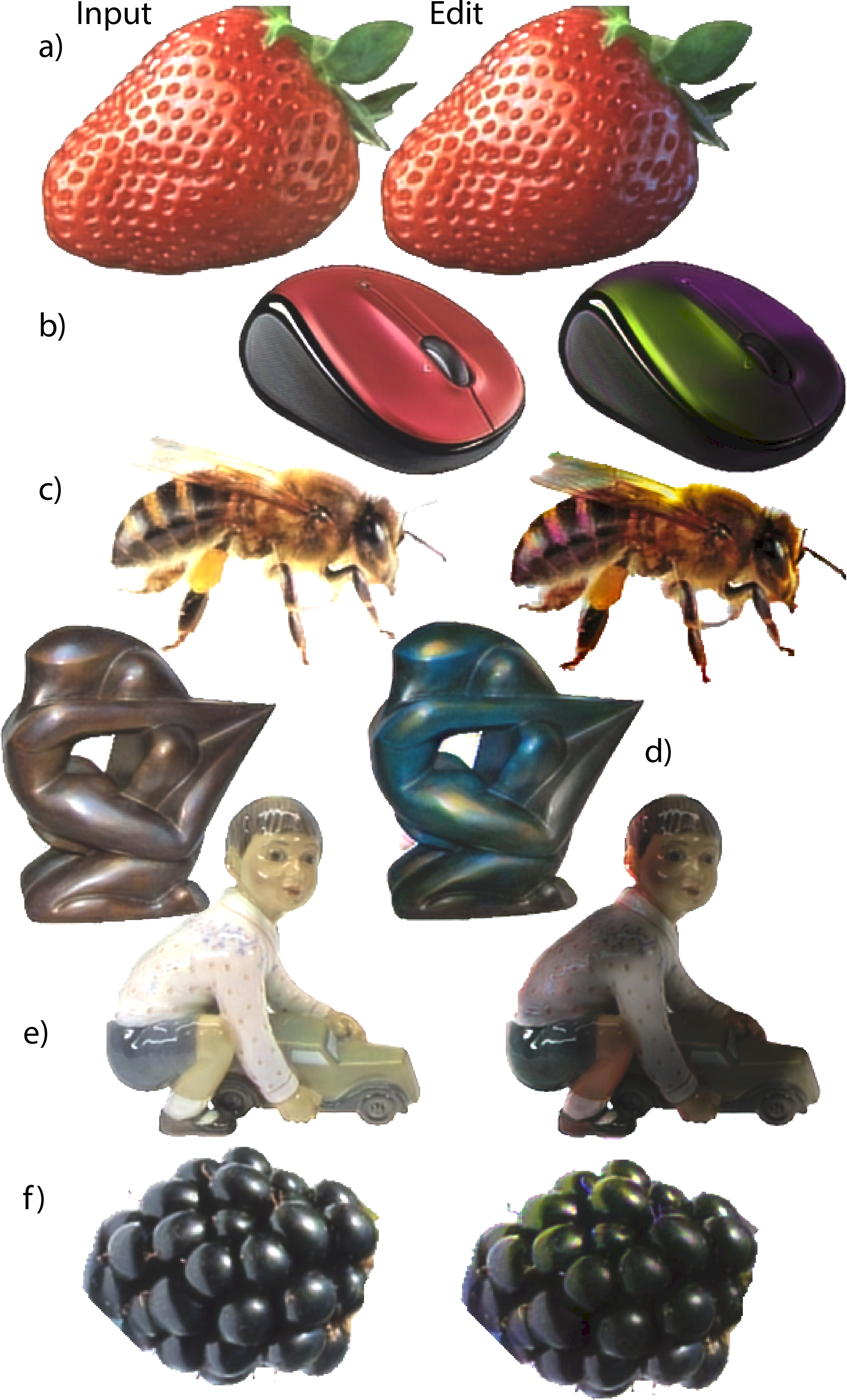}
    \caption{Directional editing (See text ``Edit'' \refSec{Results}).}
    \label{fig:DirectionalEdits}
\end{figure}

\mysubsection{Composition}{Composition}
For composition any arbitrary software that can handle layering, such as Adobe Photoshop and Adobe After Effects, can be used.
We do not limit the manipulation to produce a composition that is physically valid, because this is typically limiting the artistic freedom at this part of the pipeline \cite{todo2007locally,ritschel2009approximating,schmidt2015state}.
Our decomposition is so simple that it can be implemented using a Photoshop macro that merely  sets the appropriate additive and multiplicative blend modes, followed by a final gamma mapping.
The content is then ready to be manipulated with existing tools with WYSIWYG feedback.

\mysection{Results}{Results}

We report results in form of
typical decompositions on images,
edits enabled by this decomposition,
and numeric evaluation.
The full supplemental material with many more decompositions is found at \\\hypertarget{http://geometry.cs.ucl.ac.uk/projects/2017/PSD/results.php}{{\small{\texttt{geometry.cs.ucl.ac.uk/projects/2017/PSD/results.php}}}}.

\paragraph{Decompositions.}
How well a network performs is best seen when applying it to real images.
Regrettably, we do not know the reference light transport layer-decomposition or directional decomposition, so the quality can only be judged qualitatively.
Therefore, results of decomposing images into light transport layers is seen in \refFig{Decomposition} while decomposition into directions is show in \refFig{DirectionalDecomposition}.

\mytfigure{Decomposition}{Decomposition of input images into light transport layers. Please see ``Decomposition'' in \protect\refSec{Results}.}

\mytfigure{Comparison}{Comparison of our approach to three different reflectance and shading estimation techniques \protect\cite{BarronTPAMI2015}, \protect\cite{bell2014intrinsic}, \protect\cite{narihira2015direct}.
We run their method on real images and compare their results to ours.}

\noindent{\bf Edits.}
Typical edits are shown in \refFig{Edits} and the directional variant in \refFig{DirectionalEdits}.
Note that we support both global manipulations, such as changing the weight of all values in a layer, and local manipulations, such as blurring the highlights or albedo individually.

In \refFig{Edits}, the first car (a) change the albedo hue without affecting the highlight color.
The second car (b) removed the diffuse part resulting ins very specular car.
The banana (c) image shows increased highlights and deepened shadows.
The first shoe image (d) was made more specular and the second (e) less, while also changing albedo hue and making shadows darker.
Finally, the statue material (f) was changed to plaster by removing specular and setting albedo to identity, to bronze by removing diffuse shading and to yellow plastic by adjusting all components.

In \refFig{DirectionalEdits} the first edit (a) changed the hue of the right color to blue.
The back light on the mouse (b) was turned violet and the front light green.
The bee (c) is lit more form the side with colored light.
The bronze statue (d) is left blueish from the left.
The statue (e) is edit to be lit from the back.
The strawbery illumination (f) was made colored from the top.

\paragraph{Numerical evaluation.}

Intrinsic images assume $S$ to be zero (no specular) and combine our terms $o$ and $D$, the occlusion and the diffuse illumination, into a single ``shading'' term that is separated from the reflectance $\rho$.
\begin{align}
C = 
O  (\rho  \cdot I  + S) \approx 
O  (\rho \cdot I + 0)=
\underbrace{O\cdot I}_{\text{Shading\ }I'}\cdot\rho
\end{align}

A comparison on of our decomposition and typical approaches to generate intrinsic images is shown in \refFig{Comparison}.
In table \ref{tbl:comparison} we compare against the same techniques but on our test dataset.

\begin{table}[h]
\begin{center}
\begin{tabular}{ l c c c c }
\toprule
 Method & DSSIM & NRMSE  \\
 \midrule
 Ours & \textbf{.0661} $\rpm$ \textbf{.0146} & \textbf{.3323} $\rpm$ \textbf{.1310}\\ 
 DI & .0862 $\rpm$ .0165 & .7698 $\rpm$ .4818\\  
 IIW & .0775 $\rpm$ .0158 & .7698 $\rpm$ .4594\\
 SIRFS & .0846 $\rpm$ .0187 & 1.315 $\rpm$ 1.074\\\bottomrule
\end{tabular}
\end{center}
\caption{Evaluation of our test dataset and other intrinsic image algorithms. We report  the mean and standard deviation results of two well-known error metrics: DSSIM and NRMSE. We run the experiment on a batch of 100 examples from our test dataset comparing the grounth truth albedo to our results and our competitors'.
}
\label{tbl:comparison}
\end{table}

\paragraph{Limitations.} Like in many CNN based learning approaches, the shortcoming of our two networks are hard to pin down. Not surprisingly they perform well on training data and generalizes reasonably across other object classes, still they fail when they  see completely new type of data. While one obvious way to try to improve performance would be to add more training data (\eg different types of shape families, different illumination and materials, etc.) we would like to understand better what datasets to add to maximize improvement. This remains an elusive goal in CNN-based systems as of now.

\mysection{Conclusion}{Conclusion}
We have suggested the first decomposition of general images into light transport layers, that were previously only possible either on synthetic images, or when capturing multiple images and manipulating the scene.
We have shown that overcoming these limitations allows producing high-quality images, but it also saves capture time and removes the limitation to renderings.
Future work could investigate other decompositions such as global and direct illumination, sub-surface-scattering or directional illumination or other inputs, such as videos.

\bibliographystyle{acmsiggraph}
\bibliography{article}
\end{document}